# Investigation on domain adaptation of additive manufacturing monitoring systems to enhance digital twin reusability

Jiarui Xie, Zhuo Yang, Chun-Chun Hu, Haw-Ching Yang, Yan Lu, and Yaoyao Fiona Zhao

*Abstract*— Powder bed fusion (PBF) is an emerging metal additive manufacturing (AM) technology that enables rapid fabrication of complex geometries. However, defects such as pores and balling may occur and lead to structural unconformities, thus compromising the mechanical performance of the part. This has become a critical challenge for quality assurance as the nature of some defects is stochastic during the process and invisible from the exterior. To address this issue, digital twin (DT) using machine learning (ML)-based modeling can be deployed for AM process monitoring and control. Melt pool is one of the most commonly observed physical phenomena for process monitoring, usually by high-speed cameras. Once labeled and preprocessed, the melt pool images are used to train ML-based models for DT applications such as process anomaly detection and print quality evaluation. Nonetheless, the reusability of DTs is restricted due to the wide variability of AM settings, including AM machines and monitoring instruments. The performance of the ML models trained using the dataset collected from one setting is usually compromised when applied to other settings. This paper proposes a knowledge transfer pipeline between different AM settings to enhance the reusability of AM DTs. The source and target datasets are collected from the National Institute of Standards and Technology and National Cheng Kung University with different cameras, materials, AM machines, and process parameters. The proposed pipeline consists of four steps: data preprocessing, data augmentation, domain alignment, and decision alignment. Compared with the model trained only using the source dataset, this pipeline increased the melt pool anomaly detection accuracy by 31% without any labeled training data from the target dataset.

Keywords: Additive manufacturing, machine learning, digital twin, knowledge transfer, domain adaptation.

## I. INTRODUCTION

In-situ monitoring in additive manufacturing (AM) allows for the collection of critical in-process data during a build process. Complex physical phenomena, such as melting and cooling, can be observed by various sensors. This in-situ data provides valuable insights into build quality and may enable real-time control. Pyrometers and coaxial cameras are two popular in-situ AM sensors embedded in powder bed fusion (PBF) machines. Pyrometer is a broad term describing sensor devices that utilize the principles of pyrometry to measure temperature [1]. For example, photodiodes can proportionally record the intensity of light, providing critical information about the thermal conditions of melting [2]. Coaxial cameras operating coaxially with the laser beam can directly record the melt pool down to the micrometer level [3]. Coaxial images can provide various details about the melt pool, including its intensity, shape, and size.

In-process data acquired using the monitoring systems are valuable assets for creating process digital twins (DTs) [4]. Machine learning (ML) models, trained using historical data from experiments or simulations, can make predictions based on the in-process data for various AM tasks, including defect detection, quality prediction, and process control [5]. For example, melt pools monitored using cameras can be used to build process anomaly detection models to identify the locations of potential defects [6]. Defects on AM layer-wise images can be labeled to train defect segmentation models that classify defects at a pixel level. Combined with process parameters and part geometry information, in-process data can infer the print quality such as the geometric conformity and mechanical properties [5]. With a feedback control loop, the optimal process parameters can be adjusted in-process to mitigate defects and enhance the print quality.

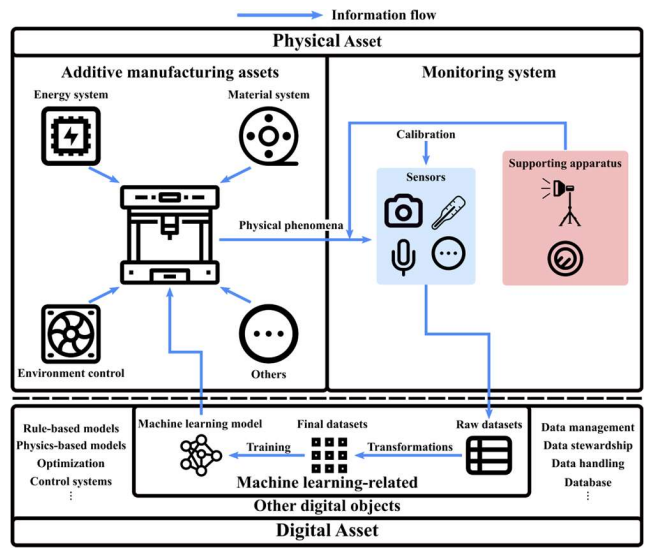

Figure 1: Schematics of an additive manufacturing digital twin.

<>Jiarui Xie received funding from the McGill Engineering Doctoral Award (MEDA) fellowship of the Faculty of Engineering at McGill University and Mitacs Accelerate Program (Grant # IT13369).

Jiarui Xie and Yaoyao Fiona Zhao are with McGill University, Montreal, Quebec, Canada (e-mails: jiarui.xie@mail.mcgill.ca and yaoyao.zhao@mcgill.ca).

Zhuo Yang is with Georgetown University, Washington, D.C., USA (email: zy253@georgetown.edu).

Chun-Chun Hu is with National Cheng Kung University, Tainan, Taiwan (R.O.C.) (chunchunhu@imrc.ncku.edu.tw).

Haw-Ching Yang is with National Kaohsiung University of Science and Technology, Kaohsiung City, Taiwan (R.O.C.) (email: hao@nkust.edu.tw).

Yan Lu is with the National Institute of Standards and Technology, Gaithersburg, MD, USA (e-mail: yan.lu@ nist.gov).

DT, a method integrating modeling and data analytics, has become increasingly relevant to manufacturing system design, development, and operation. DT centers on dynamic and virtual representations of a physical system, which synchronizes with data collected using sensors and inspections, to enable closed-loop control and/or provide optimized solutions for decision-makers [7]. With the bi-directional information flows between physical systems and their virtual twins, DTs are applied to AM to continuously mirror the AM part development process through the lifecycle. They employ multi-level and multi-fidelity models and apply advanced reasoning and fusion for various applications including process monitoring, part quality prediction, and supply chain decisions. DT is "fit-for-purpose digital representation of an observable manufacturing element with synchronization between the elements and their digital representation", defined by ISO 23247. AM DT studies have been reported for various purposes (e.g., design [8], process monitoring and control [4], and part qualification [9]).

Although AM DTs have been established and reported in numerous publications, their low reusability remains a challenge when deploying them in an altered physical asset. The physical asset of AM DT employs high variability concerning different AM processes, machines, materials, process parameters, sensor types, supporting apparatus, and sensor calibration and settings (Fig. 1). Any alteration in the physical asset will lead to domain shifts that change the statistical distributions of the raw and final datasets [10]. Therefore, the performance of the ML models trained to capture the original distributions is compromised, yielding discrepancies between the physical and digital assets. Besides, it is infeasible to gather data and train ML models from scratch for any alteration in the physical asset. To address this, knowledge transfer can be conducted from the original asset to the altered asset.

There are several categories of knowledge transfer, including transfer learning (TL), domain adaptation (DA), sequential learning, and meta-learning [11]. DA can address domain shifts and has been extensively utilized in ML-based AM process monitoring [12]. It is reported that knowledge transfer has successfully adapted source physical assets to target assets with altered AM processes, machines, process parameters, geometries, and materials. However, the existing DA publications only investigated a limited number of concurrent alteration types while keeping the rest of the physical assets the same between the source and target. In reality, a new manufacturer usually aims to reproduce a product with different machines and materials, which also leads to different process parameters. To massively deploy monitoring systems in production, manufacturers would prefer low-budget sensors, which may result in different data quality, settings, and calibration. This paper investigates and proposes a pipeline that conducts DA from an established source physical asset to a newly developed target physical asset. The source asset is the Additive Manufacturing Metrology Testbed (AMMT) at the National Institute of Standards and Technology (NIST). The monitoring system and data of the source asset are well-calibrated and registered.

The target asset is a newly deployed PBF machine and monitoring system at National Cheng Kung University (NCKU). Two datasets were acquired from the two physical assets that employ different AM machines, process parameters, materials, and monitoring systems. A labeled dataset can be obtained from the source domain, but there is no labeled training data for the target domain. The contributions of this paper are:

- Investigation on DA between different vision-based monitoring systems for PBF melt pools.
- Establishment of a DA pipeline consisting of data preparation, data augmentation, domain alignment, and decision alignment.
- Such an unsupervised DA pipeline that requires no labeled training data significantly reduces the engineering effort when reusing DTs on new AM physical assets.
- Implementation of a case study that represents an extreme domain shift scenario.

The remainder of this paper is organized as follows. Section 2 reviews the related works and highlights the research gaps. Section 3 elaborates on the methodology of the DA pipeline. Section 4 first illustrates the experiment settings of the source and target assets, then presents the results of DA. Section 5 discusses the results with respect to the distribution alignment effect and the impact of domain knowledge. Section 6 highlights the remarks of this research.

## II. BACKGROUND

This section first introduces the theories and methods of DA. Thereafter, the research works of DA in ML-based AM modeling are reviewed and summarized.

### A. Domain Adaptation

DA is a knowledge transfer strategy that transfers knowledge between a source and a target [11]. In DA, the source and target datasets have the same output variables but different input distributions. It means that the feature spaces and/or the marginal probability distributions of the two domains are different. The target domain usually suffers from small data sizes or a lack of labeled data. The two domains are assumed to be similar enough that the knowledge learned from the data-rich source domain can improve the generalization of the target model.

DA can be categorized into instance-based, feature-based, and model-based methods. Instance-based methods merge the two datasets with a weighting scheme applied to the source dataset [13]. The weighting scheme aims to align the marginal distributions of the two datasets by upvoting the source examples that are closer to the target distribution and downvoting the ones that are disparate from the target distribution. Feature-based methods construct a common feature space where the distributions of the two datasets are well aligned using expert knowledge or ML. Such a feature space embeds salient representations of both domains and thus is useful for the prediction task of the target domain. Model-based methods usually involve pre-training and fine-tuning, where a model is pre-trained using the source dataset and then fine-tuned using the target dataset. It is assumed that the pre-trained model has learned meaningful representations for the target domain and the loss function is close to the optimum. Therefore, learning the optimal target model only requires a small target dataset.

*B. Domain Adaptation in Additive Manufacturing*

Knowledge transfer has been extensively implemented in ML-based AM modeling for various types of domain shifts, including AM processes, machines, materials, geometries, and process parameters. Knowledge transfer between different AM processes usually involves different AM machines, materials, and process parameters [12]. Safdar et al. [6] established a transferability analysis framework that decomposes AM and ML knowledge into eleven knowledge components, based on which transferability metrics are designed. In the case study, they conducted knowledge transfer from PBF to directed energy deposition (DED). Using a model-based method, the anomaly detection accuracy for DED was improved from 84% with no DA to 94% with DA. Zhu et al. [14] conducted knowledge transfer from a large fused deposition modeling (FDM) dataset to a small selective laser melting (SLM) dataset. They used a model-based soft-attention-enhanced TL to train an SLM surface quality prediction model with 97% accuracy. Only a few publications focused on knowledge transfer between different AM machines of the same AM process. Aboutaleb et al. [15] proposed an instance-based Bayesian updating framework to utilize the knowledge of the historical datasets collected from different SLM machines with different powders. With limited target data, the model achieved 99% accuracy when predicting the relative density of parts printed by a new powder and a new SLM machine. Francis [16] established a model-based DA method that transforms the source model to the target model using a total equivalent amount, which is a transformation matrix defined using expert knowledge. Their case study conducted knowledge transfer from an EOS M290 machine to a Renishaw AM 400. Numerous papers conducted knowledge transfer between AM materials [17, 18], geometries [12, 19], and process parameter ranges [20, 21]. Most of them only investigated one type of domain alteration and employed model-based pre-training and fine-tuning.

The existing DA research in ML has three major gaps: 1) Lack of investigation on domain shifts caused by the monitoring system; 2) Dependency on labeled target training data; and 3) Deviation from real production scenarios. Ren and Wang [22] is the only group that focused on the monitoring system and conducted knowledge transfer from low-definition to high-definition metrology instruments. A penalty is added to the loss function according to the difference between the source and target model parameters to align the two models. However, the existing studies have not investigated domain shifts in vision-based monitoring systems, which dominate AM process monitoring and have high variability. Besides, nearly all DA methods proposed for AM DT require labeled training data, which is not usually available in reality. Pandiyan et al. [23] conducted unsupervised DA to train a PBF defect detection model for a new powder distribution and input parameter set. An associative loss is added to the classification loss to align the two input distributions in the encoded feature space. However, this method only aligns the entire source and target distributions without class-to-class alignment. It leads to the risk that the classifier trained using the source data cannot correctly classify the target examples although the two distributions are well mixed. Additionally, real-life manufacturers are interested in the reproducibility of a product, which implies that the AM machines, monitoring systems, materials, and process parameters may all have domain shifts. Nonetheless, there has not been any research that concurrently investigated these domain shifts.

## III. METHODOLOGY

This section illustrates the context and methodology to conduct DA between two domains with different physical assets.

*A. Domain Adaptation Context*

Knowledge transfer is to be conducted from an established and well-calibrated source AM DT to target AM DT under development. The source and target physical assets with different AM machines and monitoring systems are printing the same geometry using different materials and process parameters. A labeled source dataset $D_S = \{(x_{S,1}, y_{S,1}), \ldots, (x_{S,n_S}, y_{S,n_S})\}$ with a data size of $n_S$ is obtained from the source physical assets, where $x_{S,i} \in X_S$ and $y_{S,i} \in Y_S$ are the i$^{th}$ input source example and the corresponding label, respectively. An unlabeled target dataset $D_T = \{x_{T,1}, \ldots, x_{T,n_T}\}$ with a data size of $n_T$ is obtained from the target physical assets, where $x_{T,i} \in X_T$ are the i$^{th}$ input target example. The target labels ($Y_T$) is to be predicted. Due to the change in the physical assets, the marginal probability distributions between the two domains are different ($P(X_S) \neq P(X_T)$). Changes in the vision-based monitoring system might result in different image sizes and formats, thus leading to different input feature spaces ($X_S \neq X_T$). The prediction tasks are assumed to be the same; thus, the label space is fixed ($Y_S = Y_T$).

*B. Domain Adaptation Pipeline*

This proposed pipeline is unsupervised because no training label is available from the target dataset. The common strategy of unsupervised DA is to mix and match the source and target distributions. To achieve this, the pipeline consists of data preparation, data augmentation, domain alignment, and decision alignment.

**Data preparation**. The source domain is often well-calibrated and well-registered thus yielding better data quality. The target domain is usually under development with lower data quality and quantity. Thus, data preparation is implemented at first to align the formats and improve the data quality of the two datasets. The two image datasets might have different file formats, color modes, and image sizes. Image quality will differ due to different settings such as lighting conditions, pixel sizes, exposure times, focal lengths, and frame rates. Images of different formats are first converted to matrices for subsequent data preparation. Images of different modes are converted to the same mode. Afterward, the images are cropped or resized to equalize image sizes. Finally, the image defects, if any, should be mitigated to improve the image quality.

**Data Augmentation.** Data augmentation generates examples by partially altering existing samples or synthesizing new samples to mitigate representation bias in the dataset [24]. To adapt to a new domain, data augmentation is conducted to ensure that 1) the distribution of the source dataset becomes closer to the target distribution and 2) the target dataset can well represent the target distribution. Frequently utilized image augmentation methods include zooming, translation, flipping, rotation, and sharpening. It is important to note that data augmentation is not always label-preserving, which means that the labels of the generated data might differ from the original data. Therefore, data augmentation should be carefully implemented and verified to ensure the data quality.

To approximate the target distribution with the source dataset, data augmentation must be designed according to the differences between the source and target physical assets and settings. For example, the sizes of the source melt pools should be adjusted to approximate the sizes of the target melt pools. The range of the zooming ratio ($R_Z$) for the source dataset can be computed based on the pixel sizes of the source ($S_S$) and target ($S_T$) monitoring systems:

$$\left\{ R_Z \in \mathbb{R} \middle| (1 - f_Z)\left(\frac{S_S}{S_T}\right) \leq R_Z \leq (1 + f_Z)\left(\frac{S_S}{S_T}\right) \right\}, \quad (1)$$

where $f_Z$ is a zooming factor that defines the upper and lower bounds of the range. Note that the above equation might not exactly align the melt pool sizes from different domains if they have different materials and process parameters.

**Domain alignment.** This step aims to learn an encoded feature space that 1) extracts salient representations for the prediction task and 2) mixes the source and target distributions. Fig. 2 demonstrates the proposed DA training method using binary melt pool anomaly detection as an example. The ML model consists of one convolutional encoder ($G_f$) that establishes an encoded space shared between domains, two task classifiers ($G_{t1}$ and $G_{t2}$) that make predictions for the label, and a domain classifier ($G_d$) that classifies between source and target inputs. $G_f$, $G_{t1}$, and $G_{t2}$ are first pre-trained using the labeled source dataset to extract representative features for anomaly detection (Fig. 2 (a)). The two task classifiers are updated using their respective task classification losses ($L_{t1}(G_{t1}(G_f(X_S)), Y_S)$ and $L_{t2}(G_{t2}(G_f(X_S)), Y_S)$). The convolutional encoder is updated using the encoder loss ($L_{enc} = L_{t1} + L_{t2}$). Thereafter, a domain classifier ($G_d$) is added to the network to align the source and target domain distributions (Fig. 2 (b)). Again, the two pre-trained task classifiers are updated using their respective task classification losses computed based on the source dataset ($L_{t1}$ and $L_{t2}$). The salient features of both the source and target input data are extracted by the convolutional encoder, and then fed into the domain classifier. The domain classifier is trained to classify whether an input example belongs to the source or target. The domain classification loss is:

$$L_d = -\log\left(1 - G_d(G_f(X_S))\right) - \log\left(G_d(G_f(X_T))\right). \quad (2)$$

The convolutional encoder is trained adversarially against the domain classifier and is updated using the penalized encoder loss:

$$L_{enc} = L_{t1} + L_{t2} - \lambda L_d, \quad (3)$$

where $\lambda$ is a trade-off factor [25]. Such an adversarial training technique forms a competition between the convolutional encoder and the domain classifier, thus constructing an encoded space that well mixes the source and target distributions.

**Decision Alignment.** As indicated in [26], such an adversarial training strategy only mixes the two distributions, while ignoring class-specific alignments. In a binary classification task, it is possible that the target normal examples are not clustered with the source normal class, and the target abnormal examples are not clustered with the source abnormal class after domain alignment. This way, the task classifiers trained to classify the source examples will not perform well when classifying the target examples. Therefore, a decision alignment step is devised to conduct class-specific alignments (Fig. 2 (c)). Both the source and target examples are fed into the convolutional encoder and then the two task classifiers. Similar to domain alignment, the source examples are used to compute the task classification losses to extract useful features for task prediction. Additionally, a discrepancy loss ($L_{dis}$) is computed based on the predictions of the two task classifiers on the target examples:

$$L_{dis} = d\left(G_{t1}\left(G_f(X_T)\right), G_{t2}\left(G_f(X_T)\right)\right), \quad (4)$$

where $d()$ is a discrepancy metric. $G_{t1}$ is updated using $L'_{t1} = L_{t1}\left(G_{t1}\left(G_f(X_S)\right), Y_S\right) + L_{dis}$. And $G_{t2}$ is updated using $L'_{t2} = L_{t2}\left(G_{t2}\left(G_f(X_S)\right), Y_S\right) + L_{dis}$. $G_f$ is again updated based on the original encoder loss ($L_{enc} = L_{t1} + L_{t2}$). This strategy aligns the predictions of the two task classifiers by fine-tuning them using the discrepancy loss. The easiest way for the task classifiers to reduce the discrepancy loss is to aggregate similar examples. Therefore, the task classifiers are adjusted to split the normal and abnormal examples because examples from the same class belong to the same distribution. Besides, the strategy tends to align the source normal class with the target normal class and align the source abnormal class with the target abnormal class. The above four steps first mix the source and target distributions, then match examples from the same class. Besides, no labeled target data are needed during the training process.

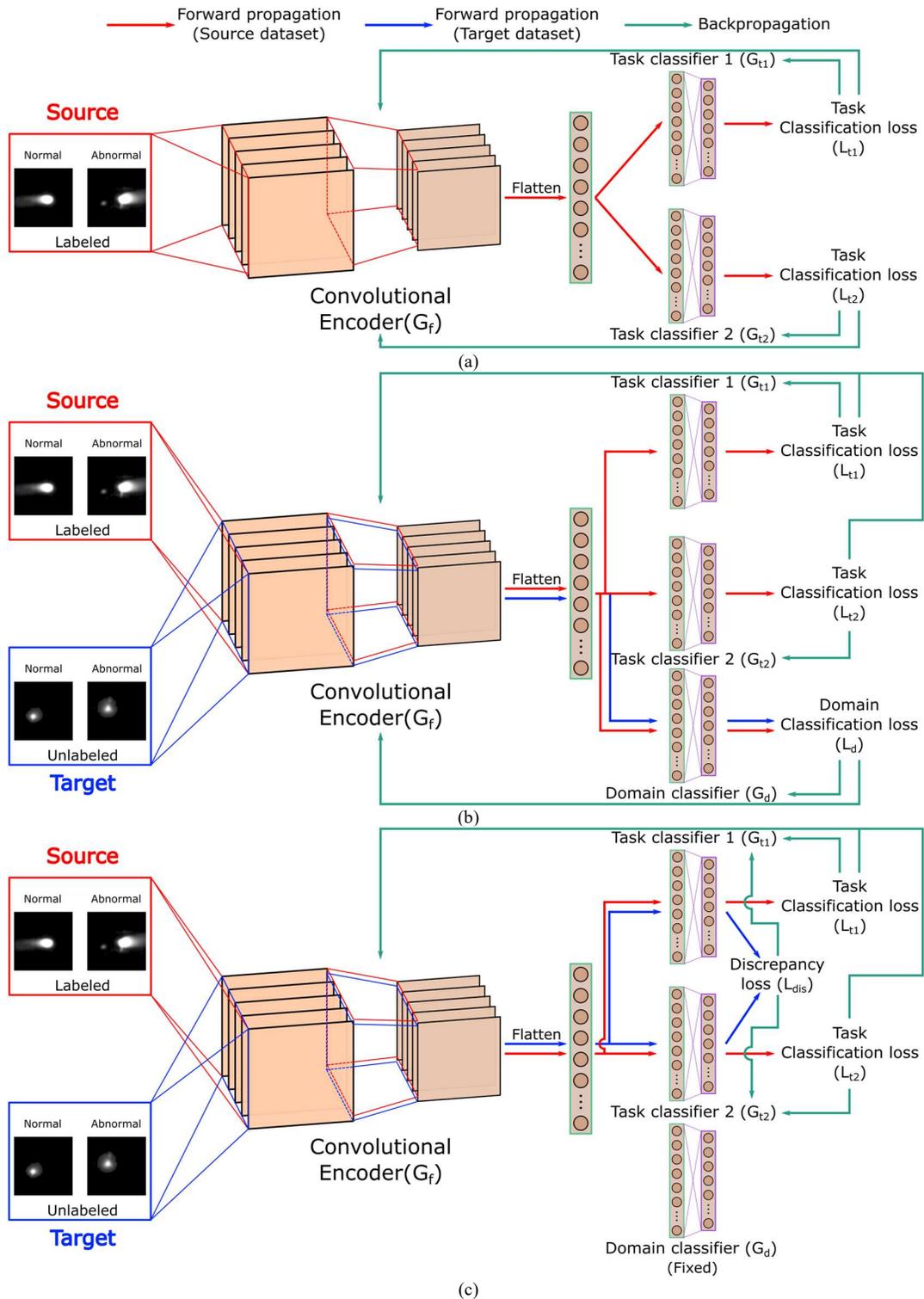

Figure 2: Training procedures of the proposed domain adaptation pipeline: a) Pre-training; b) Domain alignment; and c) Decision alignment.

## IV. RESULTS

This section illustrates the experiment settings, compares between the two datasets, and demonstrates the DA pipeline. The goal of this case study is to train an anomaly detection model using only unlabeled target training data with knowledge transfer from the source dataset.

### A. Experiment Setting

The source dataset ($D_S$) was collected from the AMMT at NIST. AMMT is an open-platform metrology instrument with capabilities of flexible control and measurement for PBF [27, 28]. It is also equipped with precise laser beam control capabilities. AMMT uses time-stepped digital commands to update laser position, power, diameter, and measurement device at every 10 µs. As a result, AMMT supports continuous laser power variation, and the monitoring signals can be fully synchronized back to the laser positions. The target dataset ($D_T$) was collected from a newly deployed monitoring system on a PBF machine at NCKU. The experiment setting of the source and target datasets are shown in Table 1. The same geometry, NIST overhang X4 part, was printed in both experiments [29].

### B. Domain Adaptation Pipeline

**Data preparation.** Melt pool images of the source and target domains are exhibited in Fig. 3. Due to different physical assets and settings, images with significantly different characteristics and data quality were obtained from two domains. The well-calibrated NIST AMMT captured consistent melt pool images with an oval melt pool and a fading tail-like plume. Minimal noise and flares were observed from the source images. The NCKU system provided less consistent melt pool images because it is under development and to be further calibrated. The melt pools have varying shapes and brightness, which might be caused by unexpected deflection. A higher level of noise and flare is observed from the NCKU images. The images were resized to 80 × 80 pixels and denoised according to their respective noise levels (Fig. 3). This DA scenario represents an extreme case of knowledge transfer where the source and target distributions are substantially different.

Table 2 shows the data split of the NIST and NCKU datasets. To mitigate data imbalance, the NIST training normal examples were downsampled to have the same data size as the abnormal examples. The NCKU training dataset employed 5819 unlabeled target images. Each validation or test set of the source and target domains was assigned 50 normal and 50 abnormal labeled images.

TABLE 1: EXPERIMENT SETTINGS OF NIST AND NCKU DATASETS.

|  | **NIST (Source)** | **NCKU (Target)** |
|---|---|---|
| AM process | PBF | PBF |
| AM machine | AMMT | Tongtai AMP-160 |
| Process parameters | **Laser**<br>Power: 195W<br>Scanning speed: 800mm/s<br>Beam size: 20µm | **Laser**<br>Power: 270W<br>Scanning speed: 600mm/s<br>Beam size: 30µm |
| Geometry | NIST<br>Overhang X4 part | NIST<br>Overhang X4 part |
| Material | IN625 | SUS420 |
| Camera settings | Image size: 120 × 120 pixels<br>Pixel size: 8x8 µm/pixel<br>Frame rate: 10,000 FPS | Image size: 160 × 160 pixels<br>Pixel size: 25 x 25 µm/pixel<br>Frame rate: 2,500 FPS |

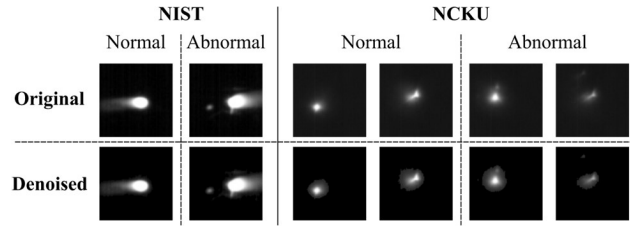

Figure 3: Comparison between the source (NIST) and target (NCKU) domain melt pool images.

TABLE 2: DATA SPLIT OF THE SOURCE AND TARGET DOMAINS.

|  | **NIST ($D_S$)** | | **NCKU ($D_T$)** | |
|---|---|---|---|---|
| *Datasets* | *Normal* | *Abnormal* | *Normal* | *Abnormal* |
| Training | 323 | 323 | 5819 (unlabeled) | |
| Validation | 50 | 50 | 50 | 50 |
| Test | 50 | 50 | 50 | 50 |

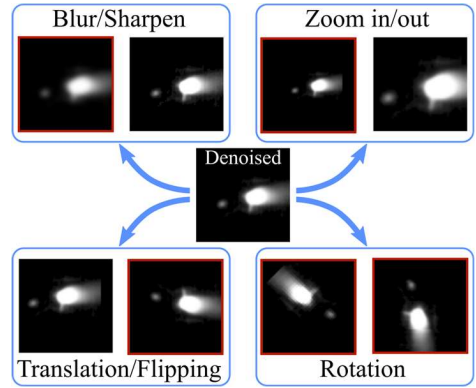

Figure 4: Visualization of the data augmentation techniques. Red boxes indicate the techniques implemented in this case study.

**Data augmentation.** Data augmentation is conducted on the source dataset to generate a larger dataset that resembles the target distribution using blurring, zooming, flipping, and rotation (Fig. 4). Instead of sharpening, blurring effects were applied to the source images to mimic the flare in the target dataset. According to (1), the source melt pools are zoomed out with a zooming ratio range of [0.3,0.35] by adding black paddings around the original images. Flipping and rotation were also implemented to approximate melt pools of different scanning directions. Translation was not chosen because it might eliminate some spatters from the images and thus unexpectedly change the labels. Ten batches of generated data with the same data size as the original dataset were obtained from data augmentation. Consequently, 6,260 synthetic training examples, 1000 synthetic validation examples, and 1000 synthetic test examples as the source examples were used to train the ML models.

**Domain alignment.** The convolutional encoder and task classifiers were first pre-trained using the source training data. A hyperparameter search was conducted to find the model with the highest source validation accuracy using Bayesian optimization. The structure and hyperparameters of the optimal convolutional encoder and task classifiers are shown in Table 3. The source test classification accuracies of the two classifiers were 92.7% and 92.4%, respectively. The loss functions of the three networks ($L_{enc}$, $L_{t1}$, and $L_{t2}$) decreased rapidly and started to converge at the 24th epoch when the pre-training phase was terminated (Fig. 5). Although the source accuracies were high, the average target test accuracy was 50% because all target examples were classified as anomalies. This is because the target distribution is significantly different from the source.

During the domain alignment phase, the domain classifier was added and trained adversarially with the convolutional encoder. Using the target validation set, a hyperparameter search was conducted to find the optimal domain classifier, whose structure and hyperparameters are shown in Table 3. There was an abrupt decrease in $L_{enc}$ at the 25th epoch because $L_d$ was subtracted from $L_{enc}$ during this phase (Fig. 5). From the 25th to the 48th epoch, $L_d$ slowly increased while $L_{enc}$ slowly decreased, indicating that the two distributions were being mixed in the encoded space. Consequently, the average validation and test accuracies gradually increased because of the alignment of the distributions. The task classification losses marginally decreased during this phase.

TABLE 3: THE OPTIMAL SETTINGS AND HYPERPARAMETERS OF THE DOMAIN ADAPTATION MODEL.

| Common settings | |
|---|---|
| Optimizer | Adam |
| Input and output dimensions | 80 × 80 and 1 |
| $L_{t1}$, $L_{t2}$, $L_d$, and $L_{dis}$ | Binary cross entropy |
| Initial learning rates | $G_f$: 1E-3; $G_{t1}$ and $G_{t2}$: 3E-6; $G_d$: 1E-5. |
| Trade-off factor (λ) | 1 |
| **Convolutional encoder ($G_f$)** | |
| *Layer* | *Hyperparameters* |
| 1st Convolutional module | Conv2d (16 channels, kernel size=3), BatchNorm2d, ReLU, MaxPool2d (kernel size=2, stride=2) |
| 2nd Convolutional module | Conv2d (32 channels, kernel size=3), BatchNorm2d, ReLU, MaxPool2d (kernel size=2, stride=2) |
| 3rd Convolutional module | Conv2d (32 channels, kernel size=3), BatchNorm2d, ReLU, MaxPool2d (kernel size=2, stride=2) |
| Flatten | Flatten (32 × 10 × 10) |
| Linear module | Linear (32 × 10 × 10, 20), BatchNorm1d, ReLU |
| **Task classifiers ($G_{t1}$ and $G_{t2}$)** | |
| *Layer* | *Hyperparameter/setting* |
| 1st Linear module | Linear (20, 32), ReLU |
| 2nd Linear module | Linear (32, 1), sigmoid |
| **Domain classifier ($G_d$)** | |
| *Layer* | *Hyperparameter/setting* |
| 1st to 3rd Linear module | Linear (20, 64), ReLU |
| 2nd Linear module | Linear (64, 1), sigmoid |

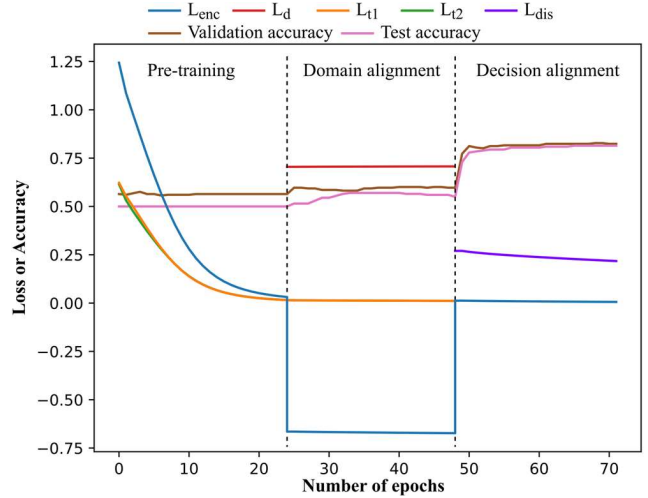

Figure 5: Loss curves of the pre-training, domain alignment, and decision alignment phases. The accuracy curves show the average accuracy of the two task classifiers on the target validation and test sets.

**Decision alignment.** In this phase, $L_{dis}$ was computed to evaluate the discrepancy between the task classifiers, whereas $L_d$ was no longer computed or subtracted from $L_{enc}$ (Fig. 5). The average target accuracies experienced a boost right after decision alignment was initiated at the 49th epoch. Afterward, $L_{enc}$, $L_{t1}$, $L_{t2}$, and $L_{dis}$ slowly decreased to match the source and target examples from the same class. The average target validation and test accuracies gradually increased and eventually reached 81.5% and 81%, respectively. Decision alignment was terminated at the 71st epoch as $L_{dis}$ started to converge.

## V. Discussion

This section discusses the DA performance, illustrates the distribution alignment effects, and investigates the impact of domain knowledge.

### A. Domain Adaptation Performance

Without any target training labels, the achievable target test performance mostly depends on the similarity between the source and target distributions. An average target test accuracy of 81% was obtained because the two datasets belong to considerably different distributions in this case study, which can be reflected by notably low target accuracies during pre-training. Although such a performance was not satisfactory for anomaly detection, it showed the effectiveness of the pipeline and helped save a lot of engineering efforts. As a comparison, labeled target training sets with different data sizes were used to train convolutional neural networks (CNNs). A hyperparameter search was conducted for each data size to find the optimal performance. It was found that 280 labeled target training examples were needed to achieve an average test accuracy of 80.5%. Besides, the model obtained from the DA pipeline became 93.5% accurate on the test set after being fine-tuned using 20 labeled target training examples. Additionally, most parts of the proposed DA pipeline can be automated. Therefore, this pipeline can save labeling efforts and extract representative features for subsequent performance improvement.

The representative features are extracted using the proposed knowledge transfer phases: domain alignment and decision alignment. As indicated in Fig. 5, the target validation and test accuracies first increased marginally during domain alignment, then drastically increased due to decision alignment. The two-step increase indicated the efficacy of the proposed phases during knowledge transfer between two considerably different distributions. Fig. 6 visualizes the source and target representations in the encoded space. The test sets were fed into the convolutional encoder to compute the encoded representations. The dimensionality of the encoded representations was reduced to 2-dimensional using t-distributed stochastic neighbor embedding (t-SNE). Without DA, the source and target distributions were disparate and thus easily differentiable. The classifier trained using the source data predicted all target examples as abnormal melt pools because they were closer to the source abnormal examples. After DA, the source and target normal examples were aggregated in the middle, whereas the source and target abnormal examples were spread away from the center. This way, the classifier learned for the source dataset can also be used to classify the target examples. However, the source and target distributions were not accurately aligned due to the huge distribution difference, leading to a target test accuracy of 81%.

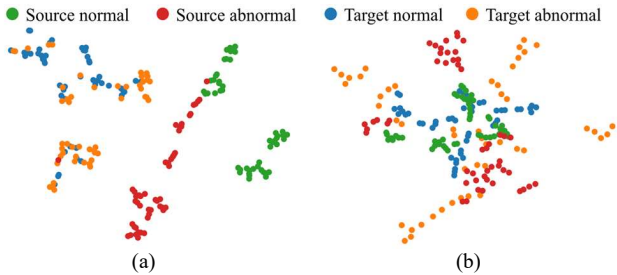

Figure 6: Visualization of the source and target test examples in the encoded space a) without DA and b) after DA.

### B. Impact of Domain Knowledge

The essence of ML is to extract knowledge from the data. However, the exploitable knowledge is limited if the target labels are not available during the learning process. Domain knowledge of the AM process and monitoring system can be utilized to improve knowledge transfer performance. In this case study, zooming data augmentation was implemented according to the zooming ratio range ([0.3,0.35]) computed based on the source and target image pixel sizes. This way, domain knowledge helped align the melt pool sizes of the two datasets to facilitate knowledge transfer. Fig. 7 reveals the change in the average target test accuracy as the zooming ratio increases. 100 models were trained using different hyperparameters for each zooming range. The models with the top 50 highest validation accuracies were selected to plot Fig. 7, which has error bars to indicate the maximum and minimum test accuracies at that range. The zooming ratio determined based on domain knowledge significantly improved the test performance of the pipeline with the highest mean and maximum test accuracies. Also, the variance of the model performance increased as the zooming ratio decreased. One potential reason is that some critical features might be removed from the melt pool images due to small zooming ratios, which introduced missing knowledge to the modeling process.

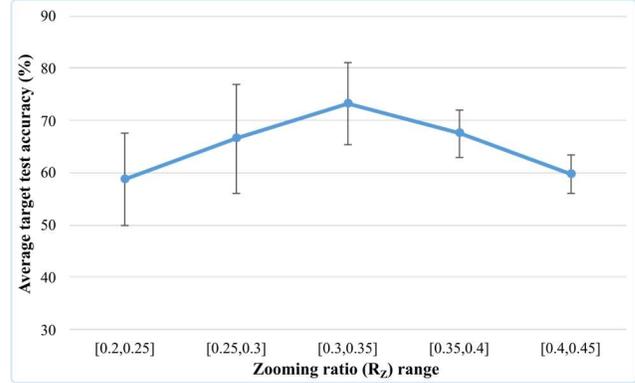

Figure 7: Impact of the zooming ratio ranges on target test accuracy.

## VI. CONCLUSIONS

This paper proposes a DA pipeline that conducts knowledge transfer between different AM physical assets to enhance the reusability of AM DTs. It is assumed that only unlabeled training data can be obtained from the target physical assets. The proposed pipeline consists of data preparation, data augmentation, domain alignment, and decision alignment. The case study conducted knowledge transfer from the NIST AMMT to a newly deployed AM asset at NCKU. Despite the substantial difference between the distributions from the two domains, the anomaly detection model achieved a test accuracy of 81% after DA without training target data needed.